# Size stability and self-agglomeration of erythrocyte-derived membrane nanovesicles versus physiological extracellular vesicles

Maryam Sanaee[*], K. Göran Ronquist, Jane M. Morrell, Muhammet S. Toprak, and Katia Gallo

Dr. M. Sanaee (ORCiDs: 0000-0001-8661-6583), Prof. M. S. Toprak (ORCID iD: 0000-0001-5678-5298), Prof. K. Gallo (ORCiDs: 0000-0001-7185-0457)

Department of Applied Physics, School of Engineering Sciences, KTH Royal Institute of technology, Stockholm 10691, Sweden

[*] Corresponding author: E-mail: msanaee@kth.se

Dr. K. G. Ronquist (ORCiDs: 0000-0001-5885-8067), Prof. J. M. Morrell (ORCiDs: 0000-0002-5245-7331)

Department of Clinical Sciences, Swedish University of Agricultural Sciences (SLU), Uppsala 75007, Sweden

**Abstract**

Extracellular vesicles (EVs) and plasma membrane-derived exosome-mimetic nanovesicles demonstrate significant potential for drug delivery. Latter synthetic provides higher throughput over physiological EVs. However they face size-stability and self-agglomeration challenges in physiological solutions to be properly characterized and addressed. Here we demonstrate a fast and high-throughput nanovesicle screening methodology relying on dynamic light scattering (DLS) complemented by atomic force microscopy (AFM) measurements, suitable for the evaluation of hydrodynamic size instabilities and aggregation effects in nanovesicle solutions under varying experimental conditions and apply it to the analysis of bio-engineered nanovesicles derived from erythrocytes as well as physiological extracellular vesicles isolated from animal seminal plasma. The synthetic vesicles exhibit a significantly higher degree of agglomeration, with only 8 % of them falling within the typical extracellular vesicle size range (30-200 nm) in their original preparation conditions. Concurrent zeta potential measurements performed on both physiological and synthetic nanovesicles yielded values in the range of -17 to -22 mV, with no apparent correlation to their agglomeration tendencies. However, mild sonication and dilution were found to be effective means to restore the portion of EVs-like nanovesicles in synthetic preparations to values of 54% and 63%, respectively, The results illustrate the capability of this DLS-AFM-based analytical method for real-time, high-throughput and quantitative assessments of agglomeration effects and size instabilities in bioengineered nanovesicle solutions, providing a powerful and easy-to-use tool to gain insights to overcome such deleterious effects and leverage the full potential of this promising biocompatible drug-delivery carriers for a broad range of pharmaceutical applications.

## 1. Introduction

Extracellular vesicles (EVs) are membrane-enclosed organellar structures released by any type of cell into the extracellular space [1], including exosomes as their smallest-size subgroup (30-200 nm diameter). Intercellular communication functionalities of EVs were first highlighted in 1982 [2, 3] with reference to prostasomes, i.e. prostate exosomes [4-6]. Since then the ability of exosome nanovesicles to interact with other cells and mediate the transfer of a variety of biomolecules triggering responses relevant for a broad range of regulatory mechanisms, immunity, disease evolution, elicited ideas of producing artificial biomimetic nanovesicles and synthetic nanoparticles with characteristics similar to EVs, as a groundbreaking platform for drug delivery and therapy [7-15].

Physiological cell-derived small extracellular vesicles and plasma-membrane-derived nanovesicles showcase remarkable features, including reduced immunogenicity and in-vivo biocompatibility [16, 17], surpassing synthetic lipid nanovesicles [18], polymer nanoparticles [19], and their metal counterparts [20]. However, to translate these membrane-derived nanovesicles into practical applications, it is imperative that they exhibit colloidal size-stability mirroring physiological exosomes, especially when stored, characterized and deployed in buffer solutions like phosphate buffer saline (PBS), a feature which is not automatically guaranteed by all preparation conditions. The nanovesicle propensity for agglomeration is an issue for which strategies such as PEGylation [21] and physiochemical modifications of buffer solutions [22] have been proposed. These approaches are however not void of deleterious side-effects, such as higher immunogenicity [23, 24] and the persistent challenges of self-agglomeration and size stability in physiological conditions pose significant hurdles and sustaining a continuing quest for appropriate solutions [25, 26], which in turn calls for the development of specialized characterization techniques capable of capturing the aggregation phenomenology in real time with a fast and robust method suitable also for high-throughput screening in solution.

Various techniques have been employed to assess the physical characteristics of exosomes and EVs-like nanovesicles, both at single-vesicle and collective levels [26-28]. Transmission electron microscopy (TEM) [29] and scanning electron microscopy (SEM) [30] are often utilized to capture images of EVs with single-vesicle resolution. However, they have drawbacks in terms low analysis throughput, labor-intensive sample preparations, time-consuming processes and most importantly, they are intrinsically destructive and not ideally suited to study vesicle properties and behavior in physiologically relevant conditions. Atomic force microscopy (AFM) [31] provides a viable alternative for imaging the morphology and size of individual nanovesicles with nanoscale precision and lends itself also to analyses in liquid solutions. Nevertheless, it has limited throughput and cannot be easily scaled to macroscopic sample assessments, which are crucial to gain a comprehensive understanding of sample size distributions and properly quantify nanovesicle aggregation likelihood and instabilities in solution. To this aim it is essential to combine single-vesicle measurement techniques with other independent macroscopic solution screening methods, which should be fast and easy to implement. Optical characterization techniques such as fluorescence correlation spectroscopy (FCS) [26, 32] and dynamic light scattering (DLS) [33-35] are particularly appealing in this regard. Traditional FCS allows for the measurement of the average hydrodynamic size at collective levels, while a newly developed FCS approach has emerged for tracking single vesicles and building appropriate statistics also on highly heterogeneous populations, as is the case of natural EVs and exosome-mimetic nanovesicles [26]. However, fluorescence microscopy techniques necessitate labeling and purification from residual free dyes in solution, a process that can be quite challenging [26]. Nanoparticle tracking analysis (NTA) [29, 36] and DLS present alternative label-free approaches, based on light scattering rather than fluorescence tagging and analysis. NTA tracks the motion of individual nanovesicles in solution by analysis of their experimental scattering data, further complemented by numerical modelling to retrieve their sizes. However, NTA typically exhibits lower reproducibility than DLS [37-39], due to the relatively short measurement trajectories of the vesicles, resulting in inherent statistical uncertainties [40, 41]. In contrast, DLS collects scattering signals from a significantly larger number of vesicles, leading to faster and statistically robust outcomes [33, 42]. Consequently, DLS stands out as a label-free, easily accessible method, demanding minimal sample volumes and enabling high-statistics determination of nanoparticle hydrodynamic sizes in solution. As a result, it has found widespread use in characterizing the size and zeta potential of EVs and similar nanovesicles [33, 34]. Nevertheless, it is worth noting that DLS methods, in their conventional implementation, typically yield only a general measurement of collective properties, to be interpreted and used with particular caution in the case of heterogeneous nanovesicle distributions.

In this study, a standard DLS commercial apparatus (Malvern Zeta-sizer ZS90 ) was utilized as a fast user-friendly and non-invasive system to retrieve experimental scattering data from natural and synthetic EVs. A comprehensive and comparative analysis protocol, combining DLS and AFM measurements, was then developed to specifically investigate the size-stability and self-agglomeration of nanovesicles in solution and further assess the effectiveness of counteracting measures involving sonication and dilution. The core idea of our approach relies on the consistent observation of two distinct size subpopulations emerging from the DLS measurements, in agreement with previous findings for prostasomes.[43-45] With the aid of independent AFM calibration measurements, the existence of a smaller-size (diameter $D \sim 30\text{-}200$ nm) component in the DLS data is consistently attributed to single vesicles, while the larger-size one to agglomerated vesicles. These insights were further developed into a methodology that allowed to rigorously assess the hydro-stability not only of physiological purified EVs (specifically prostasomes) but also of detergent-resistant membrane (DRM) domain vesicles synthesized from human red blood cells (RBCs), currently emerging as an extremely promising drug delivery platform[26, 46-48], which is however particularly prone to adverse vesicle self-agglomeration effects. Furthermore, the method is leveraged to investigate the impact of treatments such as mild sonication and PBS dilution on EVs and DRM-derived vesicles, quantifying reversible and irreversible agglomeration effects of particular relevance for the latter, which cannot be straightforwardly tracked to differences in the electrostatic properties of synthetic versus natural EVs, as highlighted by comparative zeta potential measurements simultaneously performed on the same samples.

In conclusion, the developed DLS analysis offers a rapid, sample-preserving method to quantitatively assess the agglomeration behavior of diverse nanovesicles in solution and assess the effectiveness of possible countermeasures. It provides an easily accessible and label-free approach to explore the influence of different physical attributes on size-stability, bridging the gap between colloidally produced nanovesicles and purified extracellular vesicles and offering a robust experimental tool to promote further advances in the field of drug delivery and therapy.

## 2. Results and discussion

Here, a comparative analysis conducted between two distinct types of prostasomes isolated from horse (referred to as H-EV) and bull (B-EV) seminal plasma, and nanovesicles synthesized from human red blood cells DRM-domains (DRM).

### 2-1. AFM size characterization

Following well-developed previous routines,[26, 46] high-resolution AFM measurements were routinely used to identify individual nanovesicles and confirm their spherical morphology and integrity. Size distribution measurements were conducted for the three distinct types of nanovesicles (H-EV, B-EV and DRM) under identical conditions. Figure 1a-c provides AFM images of the respective nanovesicle classes, illustrating their 3D-structural integrity and spherical shape. An image scan covering an area of 5 $\mu m^2$ was performed to obtain the size distribution for H-EV, B-EV and DRM vesicles, as presented in Figure 1d-f. More details about the AFM measurements and data analyses are provided in Supporting information (SI).

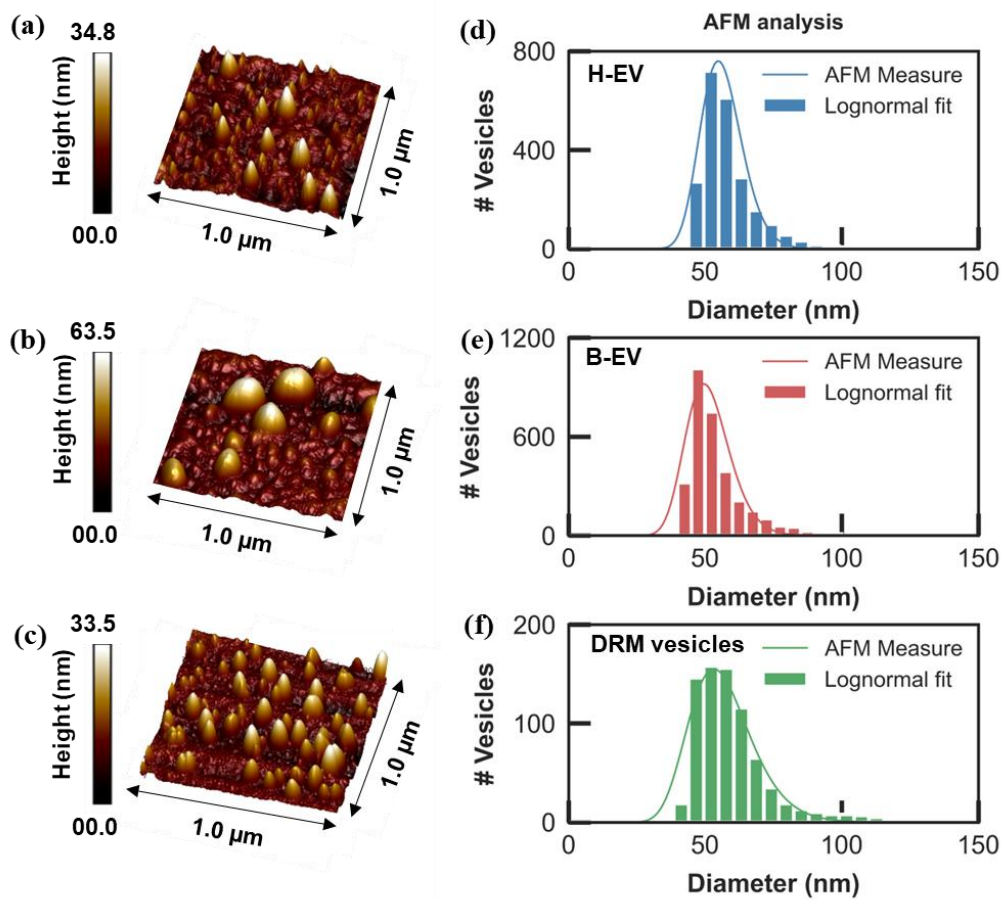

**Figure 1.** AFM images and retrieved size distributions corresponding to: a, d) horse prostasomes (H-EV); b, e) bull prostasomes (B-EV) and c, f) DRM exosome-like nanovesicles. The size distributions were obtained with a 5 nm size binning of the AFM data. The solid curves in d-f are log-normal fitting functions employed for further analysis (see also Fig. 2 and SI).

As depicted in Figure 1d-f, the size distributions of all three sets of nanovesicles exhibit similarities in both peak size and shape, demonstrating a right-skewed distribution matching the expected EV size range of 30-200 nm, typical of exosome-like nanovesicles. The observed asymmetry in the size distributions aligns well with the findings reported in various research studies through different measurement techniques such as TEM [29], FCS [26, 32], and DLS [22, 33], integrated by further modeling using dynamic scaling models [36]. Accordingly, the AFM size measurements are here fitted to a single log-normal Gaussian distribution function, represented by solid lines in Figure 1d-f. For a more comprehensive understanding, additional AFM images and detailed fits are provided in Figures S1-S3 and Table S1 within SI, respectively. It is worth noting that particular care was taken to disperse well the nanovesicles on the substrate surface and avoid agglomeration effects in the AFM measurements, as detailed in the experimental methods section and further discussed in SI-S2 (e.g. Figures S2b and S3b).

## 2-2. DLS size characterization and analysis

Applications in drug delivery and disease diagnosis impose the fundamental need to to assess the size distributions of nanovesicles and evaluate their stability in label-free manner and within physiological buffer solutions, for which DLS is an ideal solution, with rapid and easily accessibility, enabling non-invasive, large-scale collective measurements [34, 49]. Consequently, we started by conducting a qualitative comparison between the DLS size distributions and those obtained via AFM. The goal was to gain further insights and establish an analytical protocol for extracting quantitative information regarding agglomeration from the overall EV population data retrieved from the DLS measurements.

Each DLS measurement involved three consecutive sets of measurements, each lasting one minute or longer, all performed at room temperature. The Zeta-sizer software was utilized to process the raw data, resulting in three distinct datasets concerning: intensity, volume, and number distribution of nanovesicles. It is worth to note that one of the key challenges in assessing DLS data arises from the fact that the direct measurement output depends on the scattering efficiency of the nanovesicles, which scales with the sixth power of their size [35, 49] (more details provided at S3 in SI). Consequently, the peak scattering intensity from the aggregates is disproportionately represented relative to the smaller nanovesicles in the same sample, hindering an accurate assessment of the population of smaller nanovesicles [35, 49]. In contrast, the volume distribution provides a more fitting representation of the relative volume of high and low molecular weights (large-aggregates and small-individual nanovesicles), offering a more reliable size distribution to compare the relative populations of nanovesicles of different sizes [35]. A reliable conversion of the measured intensity distributions into volume distributions can be performed under the assumptions that all vesicles are spherical and possess equivalent density [35], which in our case was routinely systematically confirmed by the AFM measurements and previous extensive experience in EVs and DRM vesicle preparation (for more details see Methods).

Figure 2 presents the first set of DLS measurements made on three different nanovesicles (shaded areas). Additional DLS datasets were measured and analyzed using the same methodology and are provided in Figures S4-S6 as SI (section S3). As apparent from Figure 2a, the DLS size distribution for H-EV closely resembles the right-skewed monodispersed distribution seen in the corresponding AFM results (Figure 1d) and fits well into a single log-normal distribution. The average value of the fit residual, represented by the light blue line in Figure 2a, is less than 0.02 and deemed negligible. Meanwhile, for the two other samples (B-EV and DRM vesicles), the size distributions exhibited greater variability, with significant discrepancies arising across the Multiple sets of DLS measurements (see also data in SI, S3).

Consequently it was concluded that horse prostasomes exhibit greater stability in PBS solution, displaying no signs of agglomeration, making it a valuable qualitative reference. Based on this insight and on the AFM single-vesicle studies, the experimental data concerning the other two types of vesicles (B-EV and DRM) was analyzed by using two distinct log-normal fitting functions, attributing the lowest-size one to the single-vesicle exosome-like component of the vesicle population, in agreement with size restrictions introduced by the international society of extracellular vesicles, ISEV [50]. The specific subpopulation falling within the size range of 30-200 nm, identified as the EVs-like component (EV-Comp.), is illustrated by the solid red and green lines in Figure 2b and c, respectively. The remaining component in the DLS size-distribution data was categorized as the agglomeration component and extracted through further numerical fits (dashed lines in Figure 2b-c). Detailed information for all sets of DLS measurements and their fits can be found in Supplementary Information (section S3, Figures S4-S6 and Table S2-S4).

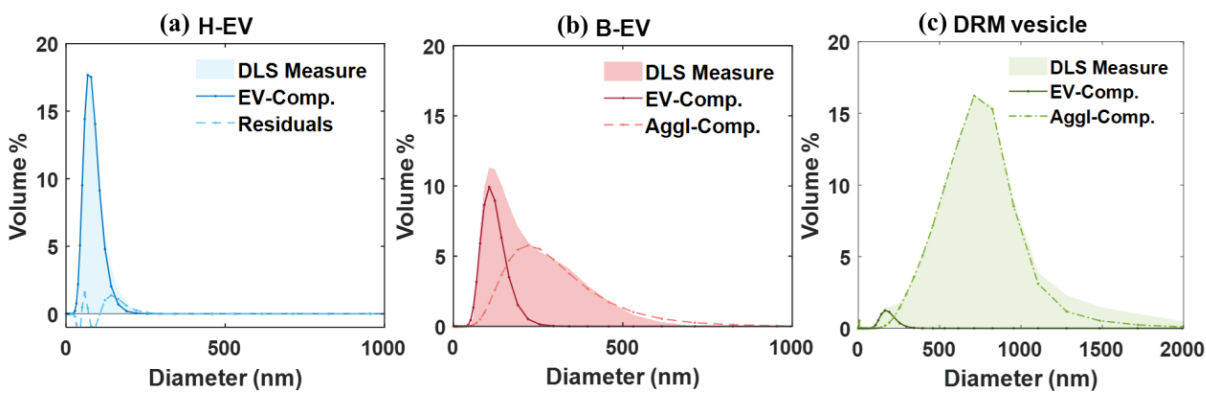

**Figure 2.** Size-distributions retrieved from DLS measurements on : a) horse (H-EV) and b) bull (B-EV) prostasomes; c) DRM vesicles. Shaded areas: DLS experimental data. Solid (EV-Comp) and dashed (Aggl-Comp) lines: EV-subpopulation and vesivle-agglomeration components, respectively, obtained from numerical fits of the experimental data. H-EV exhibit a monodispersed size distribution within the typical size range for pure EVs (30-200 nm), i.e. negligible Aggl-Comp. The analysis of B-EV and DRM vesicles reveals 52 % and 8 % EVs-like subpopulation, respectively, the rest being affected by agglomerations.

Following the above analysis protocol, the EV-Comp could be systematically isolated from each set of DLS data, and used for further quantifications and statistical studies, using a few key parameters, namely: the EVs-like average size ($EV_{ave}$), the corresponding standard deviation ($EV_{SD}$) and the area under the EV-component, normalized to the total area of the DLS distribution ($EV$ %). These metrics helped to conduct a thorough comparison and gaining deeper insights with specific concern to synthetic DRM vesicles versus natural H-EV and B-EV. Upon comparing the $EV_{ave} \pm EV_{SD}$ size, and EV % populations extracted from volume and intensity data sets, a consistent qualitative trend was observed in all cases (see also S3). As previously mentioned the volume distributions obtained from DLS data proved to be a more reliable approach to compare relative subpopulation distributions in the nanovesicle solution, although we still kept track also of the average size of the entire population (*Ave*) obtained from the direct results of overall DLS intensity measurements [35], as the main label for each data set. Figure 3 summarizes key results for natural EVs, namely H-EV (Fig. 3a-c) and B-EV (Fig. 3d-f), in terms of values of *Ave*, $EV_{ave.} \pm EV_{SD.}$, and *EV %*, obtained through repeated sets of measurements (DLS1-3). The datasets for B-EV include also results on sonicated solutions (*Sonic*), used for comparisons with DRM experiments, discussed in a later section.

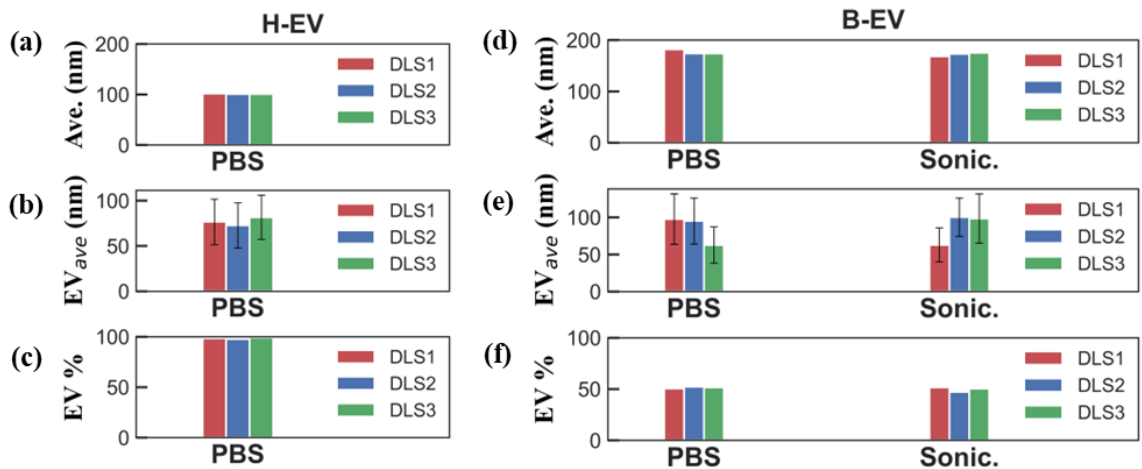

**Figure 3.** The results of DLS measurements on (a-c) horse prostasomes (H-EV) and (d-f) bull prostasomes (B-EV) before and after mild sonication. (a, d) The DLS-based Ave. size, (b, e) the estimated EVs-like average size ($EV_{ave.}$) and corresponding SD ($EV_{SD}$) as error bars, and (c, f) the EV % subpopulations are plotted for H-EV and B-EV. The H-EV showed very similar Ave. sizes around 100 nm, and they were 98 % populated at EVs-like size range. The B-EV revealed similar values for six sets of DLS measurements before and after sonication which means B-EV was stable. Meanwhile, it showed larger standard deviation compared to H-EV and only about 50-52 % of the vesicles were in EVs-like size range and the rest included microvesicles or irreversible agglomerations.

As illustrated in Figure 3a-c, H-EVs have an average size of approximately 100 nm, featuring excellent reproducibility across different measurements. More than 98 % of their population consists of purified and homogenous EVs, with a relatively small standard deviation (25 nm). In contrast, B-EVs, despite their average size falling within a reasonable range for EVs (*Ave* ~ 176 nm), present a non-negligible distribution component towards larger sizes (>200 nm) and a significantly lower EV-subpopulation component (*EV%* = 52 %), pointing out to more pronounced vesicle agglomeration effects, further investigated in the next sections. The experimental data also show more significant variations across repeated sets of measurements.

In comparison to purified natural prostasomes (H-EV and B-EV samples), the analysis of DRM vesicles revealed a much more pronounced tendency to agglomeration and instabilities in PBS solution, which we investigated with the same methodology, including further studies on the impact of sonication and solution dilution in distilled water, aspects discussed in more detail in the next sections and in SI (Tables S3 and S4).

*DRM vesicles*

Statistical studies of the average size of the overall population (*Ave*) and of the EV-subpopulation (*EV$_{ave}$*) as well as the %-contribution of the latter (*EV%*) for DRM vesicles for repeated measurements (DLS1-3) are summarized in Figure 4. The first case, labelled as *PBS* in the data sets of Fig. 4a-c, corresponds to preparation conditions identical to those adopted for the EVs of Fig. 3, i.e. DRM nanovesicles prepared with the protocols described in the methods section and dispersed in physiological PBS solution for the DLS measurements. In these initial measurements, only about 8 % of the DRM vesicles were found to be in the EV-size range. Moreover, their average DLS size exceeded by far 200 nm (*Ave* ~ 567 nm), a value significantly larger than the AFM size and well outside the typical size range of exosomes. This divergence provides strong evidence for self-agglomeration occurring of the DRM vesicles in PBS solution. However, , the effect can be counteracted by mild sonication and increased dilution (Figure 4, data labelled as *Sonic1, HalfDil* and *Sonic2*), as explained in the following sections.

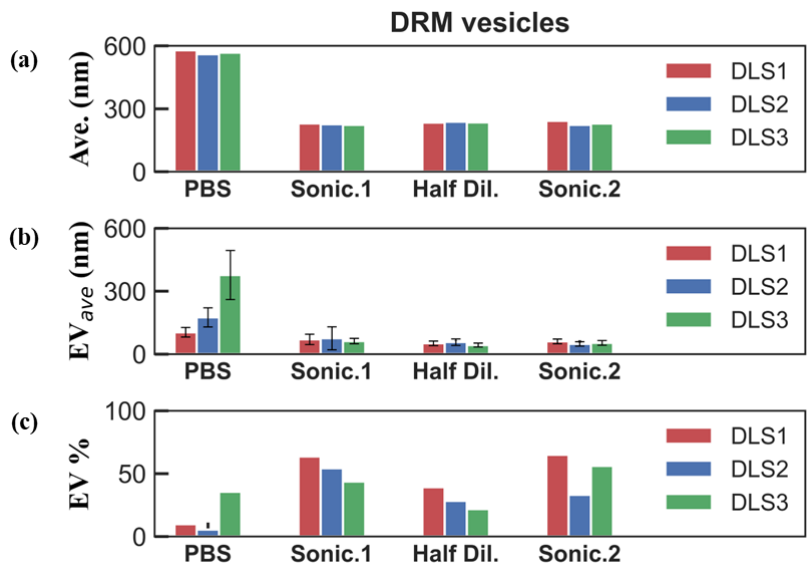

**Figure 4.** DLS studies of DRM vesicles under four different solution treatments (*PBS, Sonic1, HalfDil, Sonic2*) and repeated measurement sets (DLS1-3). Average sizes of the a) overall vesicle population (*Ave*) and b) its EVs-like subpopulation, with corresponding error bars. c) relative component of the EVs-like subpopulation of DRM vesicles in solution (*EV%*).

*Effect of mild sonication*

To evaluate the viability of mitigation measures counteracting the observed agglomeration and instabilities of the initial PBS preparations of B-EV and DRM vesicles, these samples were subjected separately to mild sonication in a water bath with a 30 % power duty-cycle at room temperature for 20 min. Subsequently, the samples underwent repeated DLS measurements. The data from the three sets of DLS measurements for B-EVs before and after mild sonication, were fitted to a log-normal distribution to extract the corresponding

EVs-like and agglomeration components (see S3 for detailed information). The B-EV results are presented under the labels *PBS* and *Sonic*, respectively, in Figure 3d-f. As depicted in these bar plots, the *Ave* and $EV_{ave.}$ of B-EV consistently hovered around 98 (177) nm and 85 (78) nm, after (before) sonication, respectively. All size measurements fell within a reasonable range for *EV %*, indicating stability of the B-EV. Nevertheless, their $EV_{SD}$ generally exceeded that of H-EV, and their *EV %* value, in line with the ISEV standard, consistently amounted to 52 % and 50 % before and after sonication, respectively. This suggested that half of the B-EV population remained distributed in larger sizes than typical EVs, which can in principle be attributed to larger extracellular vesicles, irreversible agglomerations, or a combination of both.

Figure 4a-c presents the compiled results of all sets of DLS measurements and preparation conditions considered for DRM vesicles. There the label *Sonic1* refers to DRM vesicles which underwent mild sonication under identical conditions as the B-EV of Fig. 3 (*Sonic* data), followed by DLS measurements. As apparent from Figure 4 (*Sonic1* data), a first sonication of the DRM vesicles successfully reduces $EV_{ave}$ from 140 nm to 70 nm, well within the typical size range of EVs. The sonicated samples also display lower standard deviation and a marked increase of the EVs-like component, from 8% (*PBS*) to 54 % (*Sonic1*). Hence, these findings reveal that approximately 46 % of the DRM vesicles initially display weak and reversible self-agglomeration which can be removed by mild sonication. Meanwhile, the remaining 46 % of the DRM vesicles' subpopulation (referred to as *Aggl-Comp* in Figure 3) exhibits sizes exceeding 200 nm, even after this first sonication. These larger vesicles may potentially include giant unilamellar vesicles (GUV), akin to liposomes, or be due to irreversible self-agglomeration, or correspond to a combination of both effects. In any case, the weak forces underpinning DRM vesicle agglomeration in solution can be to a good extent overcome by sonication, shifting the size distributions retrieved by DLS closer to those obtained by AFM. It is however worth to note that the specific preparation conditions required for the AFM, involving several dilutions of the original vesicle solutions in distilled water , followed by their dispersion and drying over silicon surfaces prior to the AFM measurements, essentially preclude any possibility to analyze and properly quantify agglomeration effects as well as potential countermeasures in solution.

### *Effect of PBS dilution and consequent second mild sonication*

Given the persistence of a residual large-vesicle component in the DLS data for DRM vesicles, even after their sonication, we set to investigate the impact of the ionic strength of the solution on their agglomeration, To this aim, DRM vesicles in PBS were diluted 1:2 in distilled water and subsequently subject to more DLS measurements, the results of which are presented under the label *HalfDil* in Figure 4. The dilution led to a slight increase in the overall average size of the DRM vesicles (Δ*Ave* ≈ 9.5 nm) . However, this was accompanied by a slight reduction in the average size of the EVs-like component ($\Delta EV_{ave} \approx$ -19 nm) and of the corresponding SD ($\Delta EV_{SD} \approx$ -19.2 nm). To gain further insights, such diluted samples underwent a second round of mild sonication before a fourth and last DLS measurement (data *Sonic2* in Figure 4). It was then evident that the PBS dilution followed by a second sonication had slightly altered the size distribution of the EVs-like subpopulation. The percentage of the EVs-like component of the DRM vesicles slightly increased to 63 % with an average size $EV_{ave}$ =54 nm and $EV_{SD}$ = 12 nm. This may be explained by the fact that the 1:2 PBS dilution decreases the ionic strength of the saline solution, which in turn diminishes the screening effects on the vesicles, possibly yielding more repulsive DRM vesicle-to-vesicle interactions, leading to their further dispersion. This subpopulation of the EV-component initially exhibited weak self-agglomeration, possibly induced at some point in the DRM preparation steps, involving ultracentrifugation, freezing and thawing cycles [25]. Notably, similar results were observed for DRM vesicles loaded with drugs and characterized by fluorescent microscopy in a previous report [26]. These vesicles displayed a subpopulation settling within the EVs-like size range, capable of encapsulating cargo molecules. They also exhibited a right-skewed distribution extending towards larger sizes [26]. However, the latter could not be investigated further by fluorescence methods, in contrast to the DLS study presented here.

**2-3. Zeta potential measurements**

Another crucial physical property of the nanovesicles in buffer solution is their surface charge, which can be quantified through zeta-potential measurements. This parameter might play a role in determining the stability and self-agglomeration tendency of both physiological EVs and DRM vesicles [22]. In this study, the zeta potentials of the three types of samples were measured using the Malvern system. The mean zeta potential values for the samples were approximately -17.1± 0.8 mV for H-EV, -22.0± 2.4 mV for B-EV, and -21.7± 2.3 mV for DRM vesicles. In summary, all three different samples exhibited overall relatively similar zeta potential values. However, the zeta potential values of B-EV and DRM vesicles are closer to each other (than to those of H-EV) and also have slightly larger standard deviations, which correlates weakly to the increased agglomeration tendency and further instabilities observed for B-EV and DRM vesicles and might hint to the diversity of zeta potential playing also a role in this respect.

*Polydispersity and nanovesicle self-agglomeration*

The quality of the DLS measurements was assessed using the intercept and the polydispersity index (PDI) analyses, whose results are summarized in the first and second columns of Table 1, respectively. All measurements displayed reasonable intercept values, within the reliability range of 0.1-1, with PDIs consistently below 0.5 [35].

**Table 1.** Summary of intercept, polydispersity index values and average vesicle concentration in DLS measurements for the three different samples categories (H-EV = horse prostasomes, B-EV = bull prostasomes).

| Sample | Intercept | PDI | Average concentration (vesicles/µL) |
|---|---|---|---|
| **H-EV** | 0.94 | 0.18 | $2.37 \times 10^6$ |
| **B-EV** | 0.93 | 0.24 | $4.51 \times 10^6$ |
| **DRM vesicles** | 0.95 | 0.39 | $2.57 \times 10^6$ |

Of the three different vesicle typologies, H-EV and DRM vesicles exhibited minimal and maximal PDIs, respectively. This pattern aligns with their size distributions and standard deviations. The horse prostasome displayed a predominantly pure monodispersed size distribution, containing more than 98 % EVs-like nanovesicles, resulting in the lowest PDI. In contrast, the DRM vesicles exhibited the highest level of heterogeneous size distribution, with the highest SD. This distribution consisted of a combination of GUV and self-agglomerated vesicles. Furthermore, examining the vesicle concentrations of the samples at their respective average sizes, the DRM vesicles showed comparable values to the other two, in the order of $10^6$ nanovesicles / µL. This order of magnitude corresponds to 1:100 sample dilution and agrees with previous reports on physiological EVs extracted and purified through other techniques [51-53].

## 2-4. DLS versus AFM assessments

Figure 5 summarizes the results of DLS and AFM analyses for the three different types of vesicles, by presenting the mean values of two critical parameters, namely: the average size (*Ave*) and EVs-like percentage component (*EV %*) evaluated by DLS studies (red bars), along with their error bars. Figure 5a additionally displays the average sizes resolved from AFM analyses (pink bars), for a comparison. The mean EVs populations of horse and bull prostasomes and DRM vesicles, estimated from all sets of DLS measurements, amounted to 98 %, 51 % and 37 %, respectively. As illustrated in Figure 5, the DRM vesicles display significant variability in their size and *EV %*, confirmed by their significantly larger error bars. Such a variation and limited reproducibility is to be mainly attributed to the DLS measurements performed on initial DRM sample preparations, containing a high portion of agglomerated vesicles, prior to sonication. This further highlights the importance of treatments counteracting agglomeration, in view of any characterization and utilization of the EVs-like properties of DRM vesicles in solution.

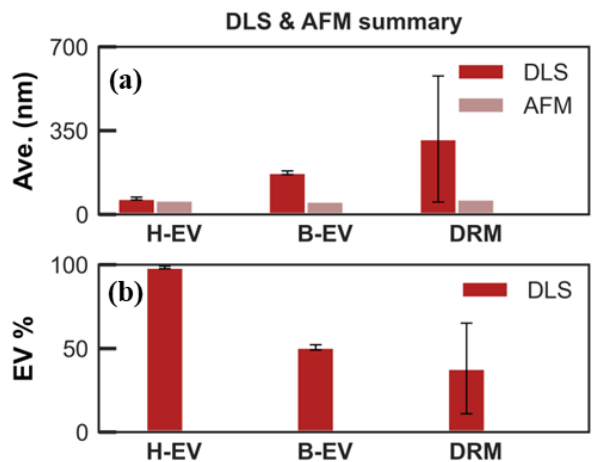

**Figure 5.** Overview of DLS and AFM results for horse (H-EV) and bull (B-EV) prostasomes and erythrocyte-derived detergent-resistant membrane vesicles (DRM). a) Average vesicle size (*Ave*) and b) EV-percentage (*EV%*) extracted from all DLS measurement statistics (red bars), with their error bars (solid lines), displayed with a) AFM-based size assessments (pink bars).

Physiological bull prostasomes (B-EV) demonstrated greater stability than DRM vesicles during mild sonication, but still displayed more agglomeration and size deviations compared to the horse prostasomes. When comparing DLS and AFM results (Figure 5a, red and pink bars, respectively) it becomes evident that the H-EV exhibit a closer size-match between DLS and AFM measurements. The AFM average size (diameter) for H-EV was $Ave^{AFM} = 59$ nm, and the difference by with the average DLS results was $\Delta D_{Ave} = Ave^{DLS} - Ave^{AFM} \sim 42$ nm, similarly to DLS-TEM assessments of exosome samples reported in Ref. [34]. The B-EV and DRM vesicles exhibited more substantial deviations in average sizes between DLS and AFM measurements, amounting to $\Delta D_{Ave} \sim 120$ and 252 nm, respectively (see Table S5 in SI for further details). As previously highlighted by the DLS investigations of sonication post-treatment, such differences can be attributed to the occurrence of both irreversible and reversible self-agglomeration in these samples. The former one is more probable for bull prostasomes, while a combination of both is observed in DRM vesicles.

## 3. Conclusion

In summary, we presented a comprehensive assessment protocol for quantitative studies on heterogeneous populations of EVs-like synthetic and natural nanovesicles, using DLS as a rapid and high-throughput measurement and analysis technique. The approach was applied to the characterization of single EVs-like and agglomerated vesicles subpopulations in purified solutions of exosomes, namely horse and bull prostasomes, as well as DRM synthetic vesicles made of human red blood cell membranes. The DLS results were further compared with AFM analysis, with good qualitative agreement for the EVs-like components of most notably horse prostasomes, but also, to a good extent, for bull prostasomes. DRM vesicles exhibited instead a much larger discrepancy between DLS and AFM size estimation results, which could be consistently traced back to spontaneous vesicle agglomeration in solution, i.e. the conditions used in DLS, as opposed to the (dried) sample preparations used in AFM. DLS then proved an unmatchingly effective analysis method to retrieve ensemble statistics on hydrodynamic size distributions, tracking both exosome-like single vesicles and agglomerated vesicle subpopulations, in physiologically and application-relevant conditions (PBS buffer solutions). A suitable model was developed and employed to evaluate size-resolved EVs-like and agglomeration-like components in DLS-based size-distributions, shedding light also on the effects of additional physical treatments such as mild sonication and PBS dilution to counteract aggregation. The DLS analyses revealed that physiological exosome types exhibited more stable EVs-like populations, amounting to 98 % and 51 % of the overall vesicle populations in PBS solution, for horse and bull prostasomes, respectively, maintaining their stability under mild sonication. In contrast, synthetic DRM vesicles displayed some reversible agglomeration and colloidal instability. Originally PBS-dispersed DRM vesicles exhibited only an 8% EVs-like content, which however increased to 54% upon mild sonication and to 63% upon solution dilution and further sonication. These findings underscore the importance of adopting suitable counteracting measures to mitigate the weak and reversible agglomeration forces occurring among natural purified exosomes, but also biomimetic exosome-like nanovesicles synthetized from human red-blood-cell membranes, in view of harnessing their disruptive potential as powerful biocompatible platforms for the therapy and diagnosis of a broad range of diseases, ranging from cancer to various neurological and cardiovascular disorders.[13, 54, 55] In conclusion, this newly-introduced DLS method offers a fast, efficient and user-friendly means to track nanovesicle properties under various preparation conditions, serving as an easy-access tool for characterizing the physical attributes of exosome-like nanovesicles in pharmaceutical applications and physiologically relevant conditions, hopefully paving the way for their deployment in large scale screening and therapeutic treatments.

## 4. Methods and experiments

The studied samples included physiological extracellular vesicles and detergent-resistant membrane nanovesicles extracted from animal seminal plasma and synthetized from human red blood cells, respectively, which were also previously studied for drug delivery applications.[26, 46] Subsequent sections provide a brief overview of the sample preparation procedures (more details can be found in Ref. [26, 46]), along with descriptions of the measurement experiments and subsequent analyses performed on these samples. More details on result assessments are provided in SI.

**4-1. Physiological EVs preparation.** Seminal plasma from horse and bull was briefly centrifuged at 2000g for 10 min. Cell debris and larger complexes were removed at 10000g and 4 ºC for 30 min. The supernatant was collected and ultracentrifuged at 100000g and 4 ºC for 1 h. The resulting pellets were resuspended in PBS and layered onto a sucrose density gradient. After preparative centrifugation at 160,000g and 4ºC for 4 h, prostasomes with a density of 1.13-1.19 g.cm$^{-3}$ were collected. Prostasomes were pelleted at 100,000g and 4ºC for 1 h, and concentrations were adjusted to 2 mg.mL$^{-1}$ using Pierce BSA protein assay kit and stored at -20 ºC until use.

**4-2. Preparation of red blood cell ghosts and red blood cell – derived DRM vesicles.** Blood bags were obtained from blood donors at Uppsala University Hospital with appropriate consent. Portions of 10 mL of red blood cells were washed in PBS (1:5) in three cycles of centrifugation at 2100g and 4ºC for 10 min. The washed RBCs were lysed in a hypotonic phosphate buffer (PB, 53.4 mOsmol.L$^{-1}$) and underwent multiple washes in PB (approximately 5-8 times 1:5) by centrifugation at 20000g and 4ºC for 30 min. The resulting RBCs ghosts were stored at -20ºC. Preparing DRM vesicles: Ten mL of the stored RBC ghosts were suspended in PBS and subjected to ultracentrifugation at 100000g and 4ºC for 1 h. The pellets were resuspended in PBS containing 1% Triton X-100 and incubated for 30min on ice. Detergent resistant membranes were floated on a sucrose density gradient with preparative centrifugation at 230000g and 4ºC for 5h. Fraction with buoyancy <1.13 g.cm$^{-3}$ was collected and identified as DRM. The DRM-fraction was pelleted at 100000g and 4ºC for 1h and the easy solubility and behavior of the pellet led us to believe that the extracted DRMs had vesiculated. The resulting pellets were resuspended in PBS and stored at -20ºC for future use.

**4-3. AFM measurements.** The vesicle samples were serially diluted (1:10) with double-distilled water, typically up to 10-12 times, to evenly disperse them onto clean silicon chips. The silicon chips (p-type) used for AFM experiments were cleaned with acetone and isopropanol, undergoing mild sonication for 5 min in each solution, and were dried with nitrogen gas. Approximately 1μL (about 11.3 ng of the sample) from each dilution was pipetted onto separate silicon chips and were allowed to dry for 3 h under a cleanroom hood at room temperature. To achieve high-resolution topographical visualization and size measurements of the vesicles, a commercial AFM system (FastScan Bruker) was employed at a slow scan rate of 0.5 Hz. AFM measurements were conducted in tapping mode under ambient air conditions using NCHV-A (Bruker) cantilevers, providing a resolution of approximately 8 nm, equivalent to the tip radius. Larger AFM images were obtained, covering an area of 5 μm² for each sample, facilitating the statistical quantification of vesicle size distributions. The resulting two-dimensional AFM images were processed in Matlab® to generate size distribution plots. More AFM images can be found at S2 in SI.

**4-4. DLS size measurements.** For DLS measurements, the vesicle samples were diluted at a ratio of 1:100 in 1 mL of PBS. These diluted samples were loaded into disposable polystyrene cuvettes. Measurements were conducted using a commercial DLS system, specifically the Malvern Zeta-sizer ZS90 (Model ZEN3690), at room temperature (25°C). The measurements were performed consecutively in three runs, each lasting longer than 60 seconds. To ensure consistency, a standard operating protocol (SOP) was established for each individual sample. This SOP included details about the refractive indexes of the PBS buffer (approximating water at ~1.33) and the vesicles (around ~1.38) [30, 31]. The ZS90 Malvern system is equipped with a red laser emitting at a wavelength of 633nm, which strikes the sample at a 90-degree angle. It captured and analyzed the fluctuations in scattered light intensity caused by the Brownian motion of nanovesicles in the solution. Utilizing the Stokes-Einstein equation, the velocity distribution of these moving vesicles was translated into their hydrodynamic diameter sizes.

**4-5. DLS data analysis.** Applying the Zeta-sizer software for data acquisition, the vesicle size distribution was presented in three different aspects: volume, intensity, and number distributions. To assess the quality of the measurements, the polydispersity index (PDI) was maintained below 0.5, and the intercept was kept below 1, serving as indicators of the reasonable quality (good) of the measurements. The size distributions obtained through volume and intensity were further analyzed to extract and compare the relative populations of different size groups. Given that the H-EV exhibited a well-defined and log-normal size distribution, it was considered as the standard. Consequently, efforts were made to identify the best-fitting log-normal distribution for the other sets of DLS size distributions. The extracted log-normal distributions within the exosome size range of 30-200 nm were considered as the EVs-like components, while the remaining distributions were associated with agglomeration for each measurement on each type of sample. For additional details regarding the fitting of all the DLS size data, refer to S3, which includes Figures S4-S6 and Tables S2-S4.

**4-6. Zeta potential measurements.** To determine the zeta potential of the vesicle samples, the same dilution (1:100) in 1 mL of PBS was employed, utilizing the ZS90 Malvern system with a dip-cell equipped with two Platinum-made electrodes. The electrophoretic mobility of negatively charged vesicles suspended in PBS was measured, and the zeta potential values were obtained using the Zeta-sizer software. Each experiment consisted of three consecutive sets of measurements, and the resulting mean and standard deviation values for each experiment were reported.

**4-7. Mild sonication.** The gentle sonication power was applied to the vesicle samples using Emmi-12HC ultrasonic in water bath system. The utilized average power was around 24 watts which was much lower than the minimum typical power in horn-tip sonicators.

AUTHOR INFORMATION

**Corresponding Author**

Maryam Sanaee, Email: msanaee@kth.se, ORCiDs: 0000-0001-8661-6583.

**Author Contributions**

M. S. and K. G. conceived the study and K. G. R. was responsible for the parts of the study that involved identity, biology, and design of the nanovesicles from three different sources and J. M. M. prepared the samples. M. S. performed the AFM measurements and their analysis. M. S. And M. S. T. performed the DLS size and zeta potential measurements. M. S. established and conducted the DLS data analysis and wrote the manuscript with contributions and approval from all authors.

 **Funding Sources**

The work was supported by the program for biological pharmaceutics of the Swedish Innovation Agency (Vinnova grant no 2017-02999), and by the OQS Research Environment for Optical Quantum Sensing of the Swedish Research Council (VR grant no 2016-06122). K. G. gratefully acknowledges further support from the Knut and Alice Wallenberg Foundation through the Wallenberg Centre for Quantum Technology (WACQT) and from the Swedish Research Council via grant VR 2018-04487.

**Acknowledgments**

We thank the Blood-donors organization in Uppsala for the blood samples.

**Declaration of interest statement.**

The authors report no competing interests.

# Supporting Information

**Size stability and self-agglomeration of erythrocyte-derived membrane nanovesicles versus physiological extracellular vesicles**

Maryam Sanaee[*], K. Göran Ronquist, Jane M. Morrell, Muhammet S. Toprak, and Katia Gallo

Dr. M. Sanaee (ORCiDs: 0000-0001-8661-6583), Prof. M. S. Toprak (ORCID iD: 0000-0001-5678-5298), Prof. K. Gallo (ORCiDs: 0000-0001-7185-0457)

Department of Applied Physics, School of Engineering Sciences, KTH Royal Institute of technology, Stockholm 10691, Sweden

[*] Corresponding author: E-mail: msanaee@kth.se

Dr. K. G. Ronquist (ORCiDs: 0000-0001-5885-8067), Prof. J. M. Morrell (ORCiDs: 0000-0002-5245-7331)

Department of Clinical Sciences, Swedish University of Agricultural Sciences (SLU), Uppsala 75007, Sweden

# Contents

## S1- Log-normal distribution function

The log-normal distribution is a right-skewed Gaussian distribution which has been well fitted to the size distribution of extracellular nanovesicles measured by AFM generally and by DLS for a specific purified extracellular vesicle type of H-EV. This distribution $G(D)$ is defined as a function of diameter ($D$) as following equation. One or the summation of a few of this type of function were used to fit and extract the EVs-like component and agglomeration subpopulations from each DLS data, while A, b and c are fitting parameters.

$$G(D) = Ae^{-(\frac{(\ln(D)-b)^2}{c^2})} \qquad (S1)$$

## S2- AFM Measurement

The vesicle morphology and size distribution were methodically examined using atomic force microscopy. To achieve this, samples containing nanovesicles in a PBS solution underwent a stepwise dilution process with distilled water, as detailed in the Methods and Experiments section within the main text. Each dried droplet's surface was then meticulously scanned using a FastScan Bruker AFM in tapping mode under ambient air conditions. A selection of representative AFM images is included in the following section.

### S2-1- AFM images

Figures S1-S3 depict comprehensive scan images of horse prostasomes (H-EV), bull prostasomes (B-EV), and DRM nanovesicles, along with detailed zoom-in images of individual vesicles. In tapping mode, the AFM recorded height and phase changes of cantilever oscillations closed to the surface. The phase images (Figures S1b, S2b, and S3b) offered valuable insights into the relative stiffness of the vesicles compared to the silicon substrate. These phase images provided clearer visualizations of the vesicles, revealing some instances of agglomerations in the case of B-EV and DRM vesicles. in the in certain areas on the surface, primarily in the case of B-EV and DRM samples. To mitigate agglomerations, we made a deliberate effort to resolve the ionic salt contents of buffer solution with several dilutions in distilled water. By clearing out the salt ions, the ionic strength of buffer and salt charge screening effects have been decreased, leading to more vesicles' repulsive interaction and less agglomeration between them. It should be kept in mind in this context that the membrane surrounding EVs is resistant to hypotonic solutions due to an unusual membrane architecture not shared with other biological membranes [1]. Therefore, most of the vesicles were showing EVs-like sizes in AFM measurements.

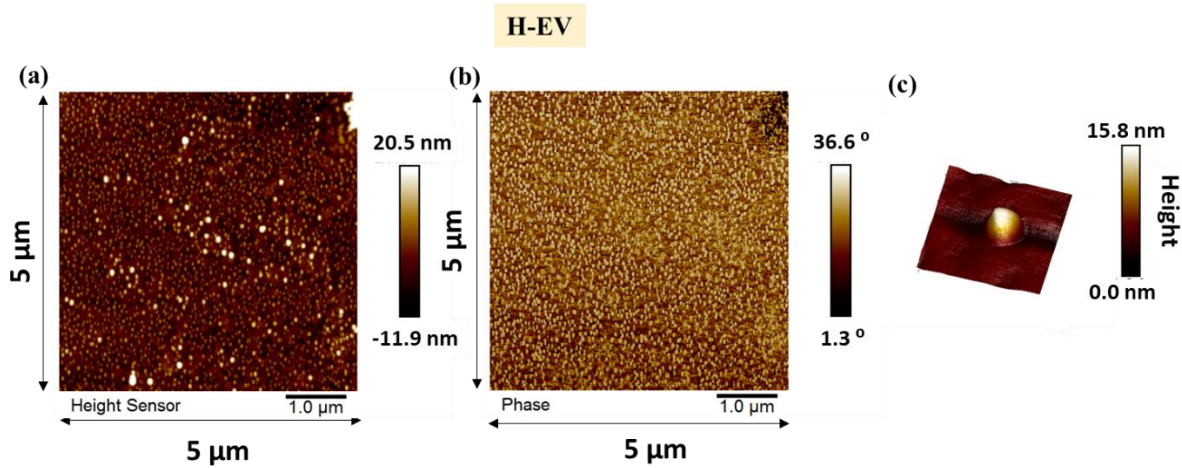

**Figure S1.** Atomic force microscopy images of horse prostasomes (H-EV) within a 5 $\mu m^2$ region, consisting of: (a) a height image, (b) a phase image, and (c) a close-up view of a typical individual vesicle.

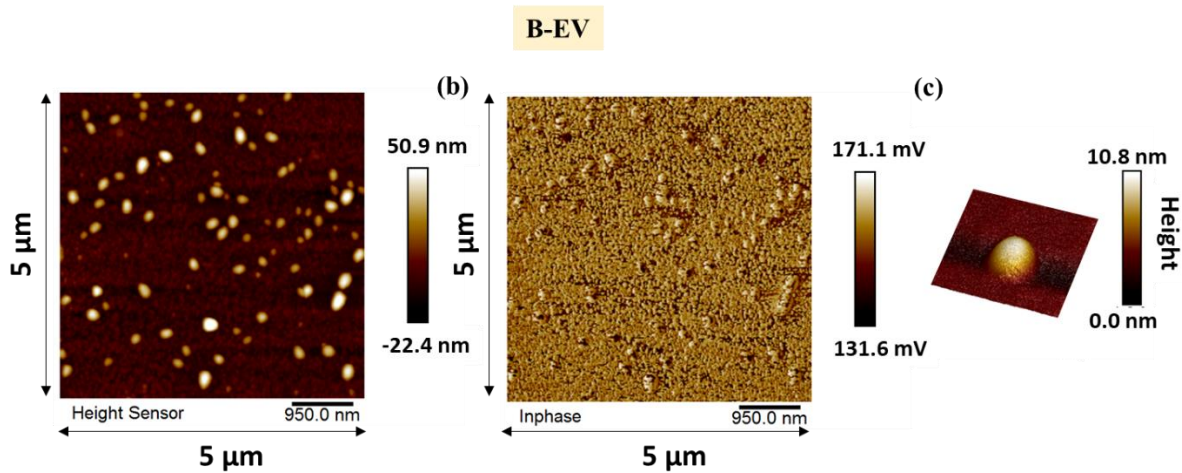

**Figure S2.** Atomic force microscopy images showcasing bull prostasomes (B-EV) within a 5 $\mu m^2$ area include: (a) a height image, (b) a phase image, and (c) a magnified view of a representative individual vesicle.

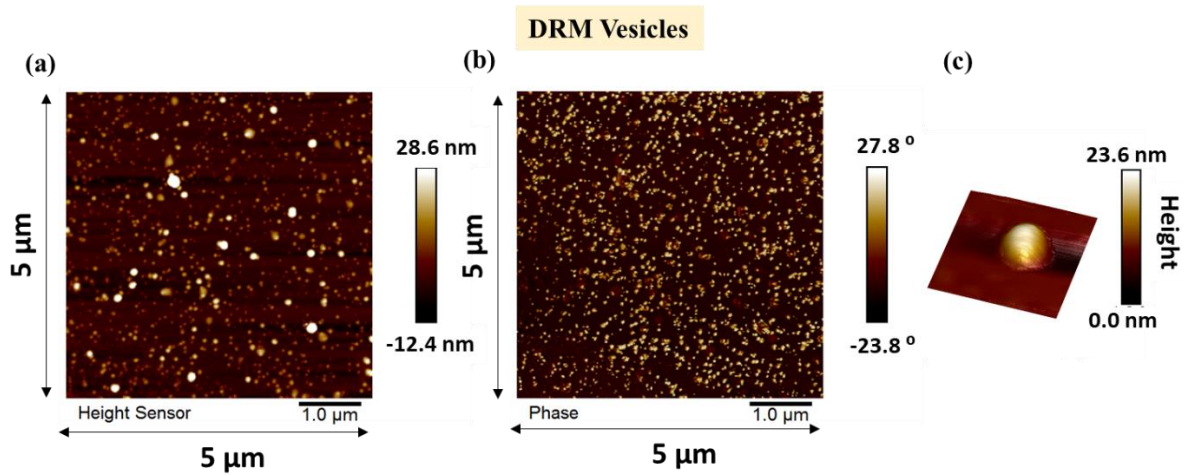

**Figure S3.** Atomic force microscopy images of DRM vesicles on area of $(5\mu m)^2$: (a) height image, (b) phase image, and (c) high resolution image of a typical single vesicle.

## 2-2- AFM Data analysis

By capturing larger AFM images with a standardized scanning area of 5 µm² for each sample featured in Figures S1-S3, we systematically quantified the size distributions of all vesicles. Utilizing image processing tools available within a commercial software package (Matlab®), we identified the nanovesicles and fitted them into circular shapes. This fitting process allowed us to extrapolate the individual vesicle diameter (d) and generate the size histograms of nanovesicles, as presented in Figure 1 of the main article. The AFM-derived size distributions were successfully modeled by a single log-normal distribution, as shown in Figure 1, and the relevant data such as the Ave. size ($D_{ave}^{AFM}$), mode size ($D_{mode}^{AFM}$) and standard deviation ($SD^{AFM}$) inferred from each AFM measurement is summarized in Table S1. Moreover, the corresponding fitted parameters are listed in this table.

**Table S1-** The Ave. size ($D_{ave}^{AFM}$), mode size ($D_{mode}^{AFM}$) and SD ($SD^{AFM}$) of three types of vesicles extracted from atomic force microscopy measurements, (H-EV = horse prostasomes, B-EV = bull prostasomes).

| | AFM analysis | | | | | |
|---|---|---|---|---|---|---|
| | Measurements | | | Fits | | |
| Sample | $D_{ave}^{AFM}$ (nm) | $D_{mode}^{AFM}$ (nm) | $SD^{AFM}$ (nm) | A | b | c² |
| H-EV | 58.74 | 53.583 | 7.145 | 760.5 | 4.004 | 0.037 |
| B-EV | 54.76 | 48.712 | 8.101 | 920.7 | 3.903 | 0.050 |
| DRM vesicles | 63.70 | 53.292 | 11.628 | 155.0 | 3.974 | 0.086 |

## S3- DLS data analysis

The DLS measurements yielded two sets of size distributions, including intensity-weighted and volume-weighted distributions for each measurement. Here, both data sets for all cases were fitted to either a single log-normal distribution or a combination of distributions. Figures S4-S6, for example, illustrate the volume data sets and their corresponding log-normal fits represented by solid lines. Detailed information, such as mode size ($EV_{mode}$), EV-average size ($EV_{ave}$), EV-standard deviation ($EV_{SD}$), and the proportion of the EVs-like component subpopulation (EV%), extracted from both volume and intensity data sets of DLS measurements, is provided in Tables S2-S4.

Practically, it becomes evident that when two distinct size species exhibit separate peak sizes, the intensity population may disproportionately represent their relative populations, given that the intensity distribution is weighted with the sixth power of the particles' sizes [2]. For example, in a simple solution containing particles of sizes $N_a$ and $N_b$, the intensity distribution of the $a$-subpopulation, denoted as $I_a\%$, can be described as [2]:

$$I_a\% = \frac{N_a a^6}{N_a a^6 + N_b b^6} \times 100 \qquad (S2)$$

Conversely, the intensity distribution can be converted to a volume-weighted distribution, which accurately portrays the relative abundance of multiple sizes, in accordance with Mie theory [2]. In line with the Rayleigh approximation, the mass of a spherical particle is directly proportional to the cube of its size, thereby allowing the volume distribution to be expressed as follows [2]:

$$V_a\% = \frac{N_a a^3}{N_a a^3 + N_b b^3} \times 100 \qquad (S3)$$

Here, $V_a\%$ represents the volume-weighted distribution, which accounts for particles of size $a$, based on their respective volumes [2]. Therefore, it was demonstrated that Volume-weighted size distributions in DRM samples depict more reliable values for the relative subpopulation of EVs-like and agglomeration components opposed to intensity, as they are highlighted in Tables S3 and S4.

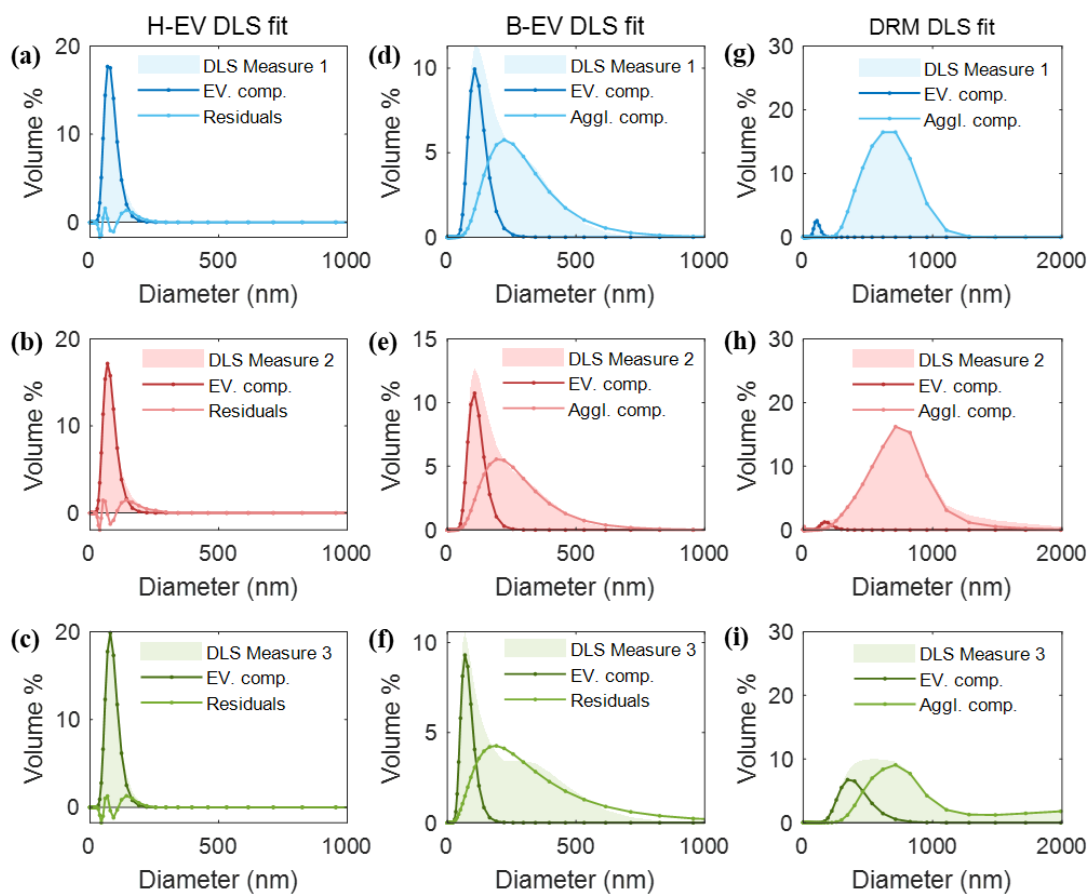

**Figure S4.** The DLS volume-weighted size distributions and extracted EV- (dark solid lines) and agglomeration components (light solid) of (a-c) horse prostasome (H-EV), (d-f) bull prostasome (B-EV) and (g-i) DRM vesicles derived from red blood cell membrane.

**Table S2**- The Ave. size and extracted information from EVs-like subpopulations ($EV_{mode}$ size, $EV_{ave}$, $EV_{SD}$, and *EV %*) for (a) H-EV: horse prostasome, (b) B-EV: bull prostasome and (c) DRM vesicles from (1) volume and (2) intensity data sets.

| DLS Volume % | | | | |
|---|---|---|---|---|
| **(a1) H-EV** | | | | |
| DLS No. | $EV_{mode}$ (nm) | $EV_{ave}$ (nm) | $EV_{SD}$ (nm) | EV % |
| 1 | 65.788 | 76.62 | 25.05 | 98.32 % |
| 2 | 61.494 | 72.64 | 24.90 | 97.63 % |
| 3 | 71.863 | 81.66 | 24.36 | 99.14 % |
| **(b1) B-EV** | | | | |
| DLS No. | $EV_{mode}$ (nm) | $EV_{ave}$ (nm) | $EV_{SD}$ (nm) | *EV %* |
| 1 | 97.747 | 111.64 | 33.98 | 50.56 % |
| 2 | 95.065 | 107.26 | 31.04 | 52.22 % |
| 3 | 62.662 | 73.38 | 24.45 | 51.66 % |
| **(c1) DRM vesicles** | | | | |
| DLS No. | $EV_{mode}$ (nm) | $EV_{ave}$ (nm) | $EV_{SD}$ (nm) | *EV %* |
| 1 | 97.104 | 104.23 | 22.92 | 9.76 % |
| 2 | 159.07 | 175.53 | 45.72 | 5.53 % |
| 3 | 328.62 | 377.3 | 117.2 | 35.54 % |

| DLS Intensity % | | | | | |
|---|---|---|---|---|---|
| **(a2) H-EV** | | | | | |
| DLS No. | Ave. size (nm) | $EV_{mode}$ (nm) | $EV_{ave}$ (nm) | $EV_{SD}$ (nm) | *EV %* |
| 1 | 101.5 | 87.75 | 108.34 | 42.08 | 99.18 % |
| 2 | 100.7 | 85.57 | 113.05 | 51.06 | 99.87 % |
| 3 | 100.8 | 90.34 | 106.14 | 35.75 | 99.99 % |
| **(b2) B-EV** | | | | | |
| DLS No. | Ave. size (nm) | $EV_{mode}$ (nm) | $EV_{ave}$ (nm) | $EV_{SD}$ (nm) | *EV %* |
| 1 | 181.9 | 142.46 | 128.10 | 82.48 | 80.77 % |
| 2 | 173.9 | 115.93 | 132.69 | 40.72 | 30.50 % |
| 3 | 174.0 | 108.77 | 160.35 | 87.14 | 45.91 % |
| **(c2) DRM vesicles** | | | | | |
| DLS No. | Ave. size (nm) | $EV_{mode}$ (nm) | $EV_{ave}$ (nm) | $EV_{SD}$ (nm) | *EV %* |
| 1 | 578.9 | 107.44 | 114.44 | 33.71 | 5.22 % |
| 2 | 559.9 | 229.30 | 348.97 | 198.36 | 22.02 % |
| 3 | 567.7 | 340.46 | 482.39 | 246.67 | 67.00 % |

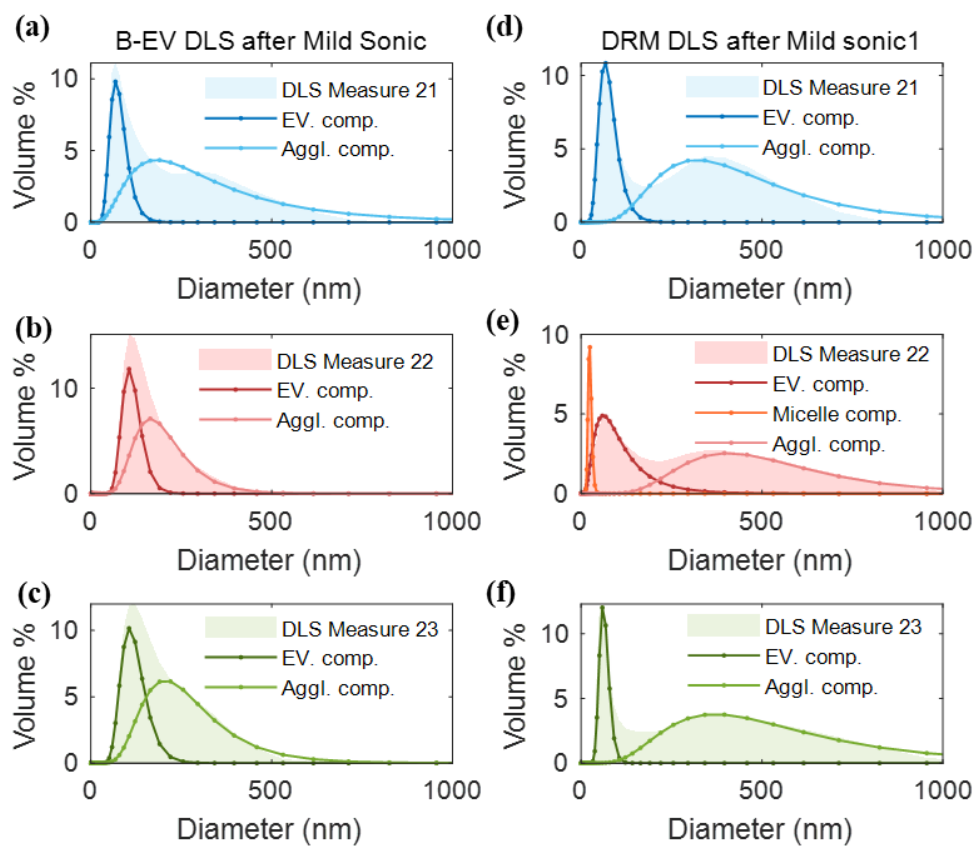

**Figure S5.** The DLS volume-based size distributions and extracted EV- (dark solid lines) and agglomeration-components (light solid) of (a-c) B-EV: bull prostasome, and (d-f) DRM vesicles after first mild sonication.

**Table S3-** The DLS results: The Ave. size and extracted information from EVs-like subpopulations ($EV_{mode}$ size, $EV_{ave}$, $EV_{SD}$, and $EV$ %) of (a) B-EV: bull prostasomes and (b) DRM vesicles after first mild sonication, extracted from (1) Volume- and (2) Intensity-weighted distributions.

| DLS Volume % | | | |
|---|---|---|---|
| (a1) B-EV *Mild Sonic.* | | | |
| DLS No. | $EV_{mode}$ (nm) | $EV_{ave}$ (nm) | $EV_{SD}$ (nm) | EV % |
| 1 | 62.822 | 72.41 | 22.82 | 51.64 % |
| 2 | 100.29 | 108.9 | 25.87 | 47.26 % |
| 3 | 71.863 | 111.63 | 33.10 | 50.43 % |
| (b1) DRM vesicles *Mild Sonic.1* | | | |
| DLS No. | $EV_{mode}$ (nm) | $EV_{ave}$ (nm) | $EV_{SD}$ (nm) | EV % |
| 1 | 58.77 | 70.08 | 24.73 | 63.60 % |
| 2 | 40.28 | 75.61 | 54.46 | 54.30 % |
| 3 | 58.35 | 62.32 | 13.21 | 43.77 % |

| DLS Intensity % | | | | |
|---|---|---|---|---|
| (a2) B-EV *Mild Sonic.* | | | | |
| DLS No. | Ave. size (nm) | $EV_{mode}$ (nm) | $EV_{ave}$ (nm) | $EV_{SD}$ (nm) | EV % |
| 1 | 102.06 | 102.06 | 145.95 | 75.74 | 54.54 % |
| 2 | 100.29 | 97.76 | 136.71 | 25.37 | 67.48 % |
| 3 | 98.367 | 126.26 | 146.62 | 47.46 | 44.05 % |
| (b2) DRM vesicles *Mild Sonic.1* | | | | |
| DLS No. | Ave. size (nm) | $EV_{mode}$ (nm) | $EV_{ave}$ (nm) | $EV_{SD}$ (nm) | EV % |
| 1 | 229.9 | 85.93 | 120.87 | 61.09 | 30.41 % |
| 2 | 225.9 | 84.39 | 128.35 | 72.82 | 22.05 % |
| 3 | 222.5 | 63.70 | 72.45 | 21.68 | 10.61 % |

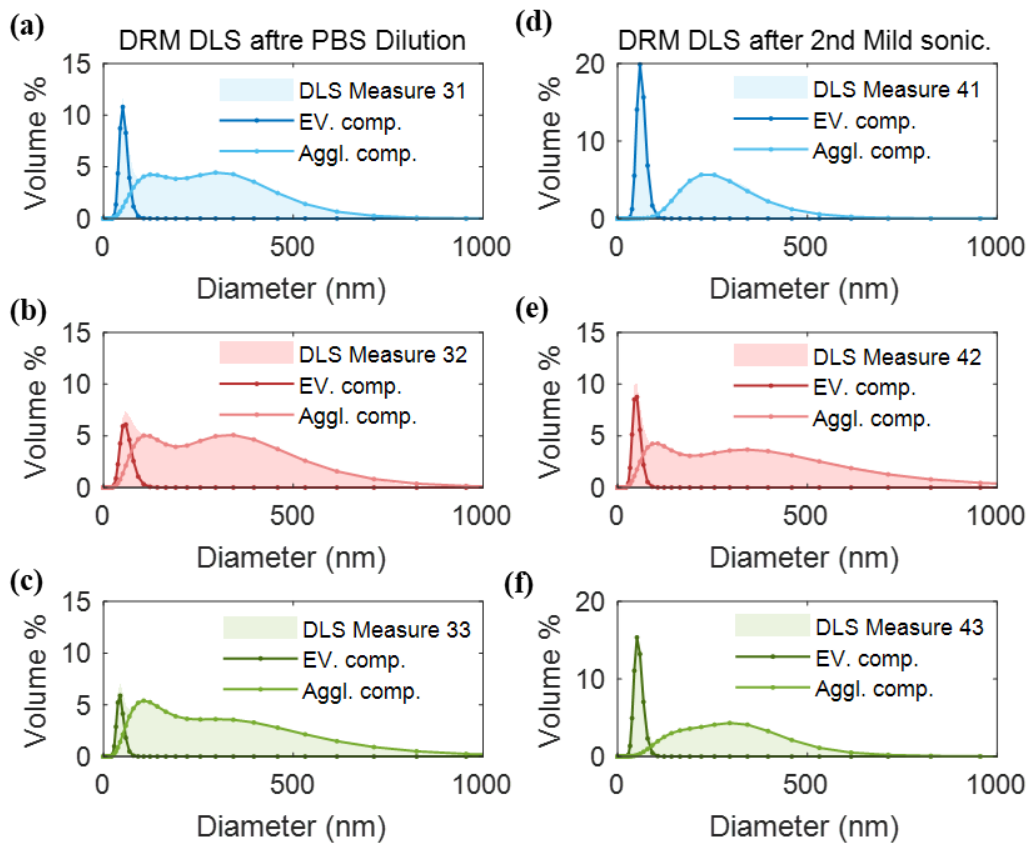

**Figure S6.** The DLS volume distributions and extracted EVs-component (dark solid lines) and agglomeration component (light solid) of DRM vesicles (a-c) after PBS half dilution and (d-f) after second mild sonication.

**Table S4-** The Ave. size and extracted information from EVs-like subpopulations ($EV_{mode}$ size, $EV_{ave}$, $EV_{SD}$, and $EV\%$) for DRM vesicles (a) after PBS dilution and (b) after second mild sonication extracted from (1) volume and (2) intensity distributions.

| DLS Volume % | | | | |
|---|---|---|---|---|
| **(a1) DRM vesicles *Half Dil.*** | | | | |
| DLS No. | $EV_{mode}$ (nm) | $EV_{ave}$ (nm) | $EV_{SD}$ (nm) | EV % |
| 1 | 48.14 | 51.49 | 11.04 | 39.19 % |
| 2 | 51.55 | 57.30 | 15.48 | 28.38 % |
| 3 | 40.39 | 43.24 | 9.33 | 21.81 % |
| **(b1) DRM vesicles *Mild Sonic.2*** | | | | |
| DLS No. | $EV_{mode}$ (nm) | $EV_{ave}$ (nm) | $EV_{SD}$ (nm) | EV % |
| 1 | 57.42 | 60.66 | 11.72 | 65.02 % |
| 2 | 45.48 | 48.64 | 10.41 | 67.54 % |
| 3 | 49.85 | 53.33 | 11.45 | 56.10 % |

| DLS Intensity % | | | | | |
|---|---|---|---|---|---|
| **(a2) DRM vesicles *Half Dil.*** | | | | | |
| DLS No. | Ave. size (nm) | $EV_{mode}$ (nm) | $EV_{ave}$ (nm) | $EV_{SD}$ (nm) | EV % |
| 1 | 233.7 | 62.51 | 92.69 | 50.80 | 11.25 % |
| 2 | 238.0 | 67.45 | 78.55 | 25.68 | 5.65 % |
| 3 | 235.1 | 46.28 | 50.28 | 11.99 | 10.35 % |
| **(b2) DRM vesicles *Mild Sonic.2*** | | | | | |
| DLS No. | Ave. size (nm) | $EV_{mode}$ (nm) | $EV_{ave}$ (nm) | $EV_{SD}$ (nm) | EV % |
| 1 | 241.7 | 57.42 | 60.66 | 11.72 | 13.45 % |
| 2 | 222.5 | 57.42 | 60.66 | 11.72 | 4.31 % |
| 3 | 229.4 | 57.42 | 60.66 | 11.72 | 7.58 % |

## S4- AFM-DLS comparison

For the purpose of comparison, the PDIs, the difference between mean values of the DLS based parameters (Ave. size, $EV_{mode}$ size, and $EV_{SD}$ obtained from EVs-like components) with their corresponding results from AFM measurements are summarized in Table S5. These parameters are defined as: $\Delta D_{Ave} = Ave^{DLS} - Ave^{AFM}$, $\Delta D_{mode} = Mode^{DLS} - Mode^{AFM}$, and $\Delta SD = EV_{SD}^{DLS} - SD^{AFM}$ in this table. It is demonstrated that the H-EV sample exhibited a smaller difference between the AFM and DLS results which was attributed to its monodespersity and hence lower polydispersity index.

**Table S5-** The mean values and according SD of zeta potential, and DLS PDIs besides the difference between AFM and DLS results on mean size, mode size and SD of three vesicle samples. (H-EV = horse prostasomes, B-EV = bull prostasomes).

| Sample | DLS PDI | $\Delta D_{Ave}$ (nm) | $\Delta D_{mode}$ (nm) | $\Delta SD$ (nm) |
|---|---|---|---|---|
| **H-EV** | 0.18 | 42.26 | 12.80 | 17.62 |
| **B-EV** | 0.24 | 119.67 | 37.45 | 20.44 |
| **DRM vesicles** | 0.39 | 251.73 | 32.18 | 17.30 |